\documentclass[10pt,conference]{IEEEtran}
\IEEEoverridecommandlockouts 

\usepackage{cite}
\usepackage{flushend}
\usepackage{amsmath,amssymb,amsfonts}

\usepackage{algorithmic}

\usepackage{graphicx}

\usepackage{multirow}

\usepackage{textcomp}

\usepackage{url}

\usepackage{pdflscape}

\usepackage[table,xcdraw]{xcolor}
\usepackage[many]{tcolorbox}

\usepackage{fontawesome}

\usepackage{hyperref}

 \usepackage{todonotes}
\newcommand{\ReviewerA}[1]{\textcolor{black}{#1}}  
\newcommand{\ReviewerB}[1]{\textcolor{black}{#1}}   
\newcommand{\ReviewerC}[1]{\textcolor{black}{#1}} 

\definecolor{main}{HTML}{CFCFCF}  
\definecolor{sub}{HTML}{CFCFCF}   

\tcbset{
    sharp corners,           
    colback = white,         
    before skip = 0.2cm,     
    after skip = 0.5cm       
}

\newtcolorbox{boxC}{
    colback = sub,  
    boxrule = 0pt   
}

\newcounter{keyTakeAwaysCounter} 
\def\BibTeX{{\rm B\kern-.05em{\sc i\kern-.025em b}\kern-.08em
    T\kern-.1667em\lower.7ex\hbox{E}\kern-.125emX}}

\AtBeginDocument{%
  \providecommand\BibTeX{{%
    Bib\TeX}}}

\begin{document}


\title{Network Centrality as a New Perspective on Microservice Architecture}


 \author{\IEEEauthorblockN{Alexander Bakhtin, Matteo Esposito,  Valentina Lenarduzzi, Davide Taibi}

 \IEEEauthorblockA{\textit{University of Oulu}, Finland}

 \IEEEauthorblockA{\{alexander.bakhtin; matteo.esposito; valentina.lenarduzzi; davide.taibi\}@oulu.fi}
 }


\maketitle

\begin{abstract}
\textit{Context}: Over the past decade, the adoption of Microservice Architecture (MSA) has led to the identification of various patterns and anti-patterns, such as Nano/Mega/Hub services. Detecting these anti-patterns often involves modeling the system as a Service Dependency Graph (SDG) and applying graph-theoretic approaches.
\textit{Aim}: While previous research has explored \ReviewerC{software metrics (SMs)} such as size, complexity, and quality for assessing MSAs, the potential of graph-specific metrics like network centrality remains largely unexplored. This study investigates whether centrality metrics (CMs) can provide new insights into MSA quality and facilitate the detection of architectural anti-patterns, complementing or extending traditional \ReviewerC{SMs}.
\textit{Method}: We analyzed 24 open-source MSA projects, reconstructing their architectures to study 53 microservices. We measured SMs and CMs for each microservice and tested their correlation to determine the relationship between these metric types.
\textit{Results and Conclusion}: \ReviewerA{Among 902 computed metric correlations, we found weak to moderate correlation in 282 cases}. These findings suggest that centrality metrics offer a novel perspective for understanding MSA properties. Specifically, ratio-based centrality metrics show promise for detecting specific anti-patterns, while subgraph centrality needs further investigation for its applicability in architectural assessments.
\end{abstract}


\section{Introduction}
Microservice Architecture (MSA) is frequently modeled as a Service Dependency (Call) graph (SDG) \cite{cerny2024static}. Recently, our community has been increasingly leveraging it for patterns or anti-patterns\cite{bakhtin2022survey, al2022using, adams2023evolution}.

Given an SDG, we can measure node centrality \cite{farsi:hal-03825330, gaidels2020service}, i.e.,  the relative importance of nodes in a network, delivering a new perspective from which recent works in Software Engineering and Architecture draw exciting results. 
Studying the relationship, if any, between node centrality in a microservice SDG and \ReviewerC{software metrics (SMs)} can uncover valuable insights into system architecture, performance, and maintainability \cite{cerny2024static,10.1007/978-3-319-92901-9_8 }. Central nodes often play critical roles in communication and computation, making them key to identifying bottlenecks, scalability challenges, or potential points of failure. Correlating centrality with metrics like size, complexity, and quality can improve optimization, testing, and refactoring efforts \cite{esposito2023uncovering, findik_2024}. 

In fault localization tasks specific to cloud architecture, Mariani et al. \cite{mariani2018localizing} utilized centrality algorithms as part of their LOUD framework on the propagation graph of KPIs to identify the failing resources. \ReviewerB{Wu et al. \cite{9527007} adopted a dynamically reconstructed SDG to build a metric causality graph, on which they calculate the centrality scores to determine the root causes of anomalies. Ma et al. \cite{8818432} leveraged a similar approach, but they omitted the SDG generation step.} Li et al. and d'Aragona et al.~\cite{li2023analyzing, li2023evaluating} have used centrality algorithms extensively on the network of developer collaboration in microservice development to identify core contributors and key developers.

However, to our knowledge, no previous work has investigated how the diverse centrality metrics of an MS are related to other classical approaches for measuring the overall quality and ``importance'' of a software entity, usually obtained via static code analysis (SCA).
Hence, we designed this work to provide a preliminary answer and pave the way for a more in-depth analysis.

We employed SCA tools to gather SMs such as size, complexity, and quality for 53 microservice from 13 systems, reconstruct their architecture, and study the relationship of different metrics with microservice centrality. Thus, our work provides the following contributions: (\textbf{i}) we provide a unified dataset containing Java Spring OSS microservice projects, their architecture, and microservice-level SMs; (\textbf{ii}) we perform the first investigation on the relationship of SMs with node CMs.

Our findings suggest that microservice CMs are weak to moderately correlated, when statistically significant, with traditional SMs, except when the correlations arise from the SDG's structural properties. Finally, we highlight a chance for centrality metrics to fundamentally provide another perspective for microservice architecture analysis. For instance, centrality can be used to identify anti-patterns such as Hub-like, Nano, and Mega-services. 


\textbf{Paper structure}:
Section \ref{sec:back} presents the background of the static analysis techniques, Section \ref{sec:related} discusses the related works, Section \ref{sec:method} presents the empirical study design, \ReviewerC{Section \ref{sec:result} presents our findings and Section \ref{sec:discussion} discusses them. Finally, Section \ref{sec:threats} addresses the threats to validity of this work and Section \ref{sec:conclusion} concludes the paper.}

\section{Background}
\label{sec:back}
In this section, we describe the background concerning architectural reconstruction using static analysis and calculation of size, complexity, quality, and centrality metrics.

\subsection{Static architecture reconstruction}

Static analysis refers to the analysis of the source code (in certain cases - bytecode \cite{10589728,esposito2024extensive}) of a system to gain insight into the system's design and architecture, code quality, and presence of patterns/anti-patterns/smells \cite{bakhtin2022survey,esposito2024validate}. Bakhtin et al.~\cite{bakhtin2023tools} performed a Systematic Mapping Study of static analysis tools aiming in particular at reconstructing microservice architecture. Together with additional authors, they currently attempt to run the tools and consider their effectiveness in achieving this task \cite{schneider2024comparisonrr, schneider2024comparison}.

Our study used the \texttt{Code2DFD} tool by Schneider et al.~\cite{schneider2023automatic}, \ReviewerB{which is identified in \cite{schneider2024comparison} as one of the most effective and generalizable tools.} This tool targets exclusively Java projects leveraging the Java Spring Boot framework for building microservices.

\subsection{Size metrics}
We use the \texttt{Understand} tool by SciTools\footnote{\url{https://scitools.com/}} to compute the size metrics of a project.
\texttt{Understand} is a commercial tool that is adopted in the industry\footnote{The tool's website mentions companies like Dell, IBM, and Samsung}. It is thus a convincing choice for practitioners facilitating the potential adoption of the proposed analysis in the industry while also providing free licenses for educational and academic use, thus enabling other researchers to replicate and extend our work.

Metrics provided by \texttt{Understand} include counts of lines of code, comments, total lines, number of classes, and methods. See the Online Appendix for the full list of metrics \cite{anonymous_2024_replication}.

\subsection{Complexity metrics}
\texttt{Jasome}\footnote{\url{https://github.com/rodhilton/jasome}} is a static analysis tool that can compute many different complexity metrics for the Object-Oriented paradigm in Java. 
It provides the metrics on \emph{package}, \emph{class}, and \emph{method} levels. 
The metrics provided, among others, include McCabe's Cyclomatic Complexity, McClure's Complexity, Depth of Inheritance Tree, Weighted Methods per Class.
Additionally, some of the provided metrics, such as the Number of Classes/Methods, are size metrics.
See the Online Appendix for the full list of metrics \cite{anonymous_2024_replication}.

\subsection{Quality metrics}
\texttt{SonarQube} is an industrial tool\footnote{\url{https://www.sonarsource.com/products/sonarqube/}} used for assessing the software quality. It also provides a free OSS version, which the researchers frequently use for studies related to software quality and technical debt \cite{fontana2019architectural, BALDASSARRE2020}. 
It provides metrics such as sqale index, remediation effort, ratings of security and reliability, and duplicated blocks.
A handful of provided metrics, such as cognitive complexity, are complexity metrics.
See the Online Appendix for a full list of metrics \cite{anonymous_2024_replication}.

\subsection{Centrality metrics}
In network science, centrality metrics (CMs) are defined to assess the relative importance of a node in the network due to its connections \cite{rodrigues2019network}. Different definitions of node centrality exist \cite{rodrigues2019network, borgatti2005centrality, bloch2023centrality}. \ReviewerC{Several sources from the network science community \cite{borgatti2005centrality,bloch2023centrality} guide us to the most widely used centrality scores: degree CM, betweenness CM, closeness CM, and eigenvector CM. In this work, we consider those and the following additional metrics, all accessible in \texttt{NetworkX} Python package:}
    \textit{In-degree CM} - the number of incoming calls;
    \textit{Out-degree CM} - the number of outgoing calls;
    \textit{Degree CM} - the total number of calls;
    \textit{Eigenvector CM} - values in the eigenvector of the adjacency matrix of the network;
    \textit{Betweenness CM} - the fraction of all the shortest call chains in the SDG that include a given microservice;
    \textit{Current flow betweenness CM}  - the fraction of all the random walks that pass through the node;
    \textit{Load CM} - the fraction of all the shortest call chains that pass through a given microservice;
    \textit{Closeness CM} - the average distance of the microservice to all neighbors that reach it, inverted;
    \textit{Information CM} - the average length of a random walk connecting the given service to all other reachable ones in the SDG;
    \textit{Harmonic CM} - sum of inverted distances of all the other microservices to a given microservice;
    \textit{Subgraph CM}  - the total number of all closed paths that start and end with the given microservice.


\section{Related work}
\label{sec:related}
In this section, we present the work related to gathering microservice systems for our analysis and previous work on analyzing software metrics.

We could identify only 5 relevant papers claiming to have a dataset of microservice systems: \cite{imranur2019curated, brogi2018towards, schneider2023microsecend, amoroso2024dataset, yang2024feature}. 

The earliest paper, by Brogi et al.~\cite{brogi2018towards}, presents the idea of constructing a \emph{reference} dataset of microservice projects and the authors not attempt to discover available OSS ones. The provided repository contains only a couple of reference systems with a couple of microservices each\footnote{\url{https://github.com/di-unipi-socc/microset}}.

The first attempt to catalog available OSS Microservice systems is due to Imranur et al.~\cite{imranur2019curated}.
The authors have searched GitHub repositories with term \emph{microservice}, included \texttt{Dockerfile} and language Java; looked through the first 1000 results filtering out frameworks and libraries, and only reported the top-20 results. 

Schneider et al.~\cite{schneider2023microsecend} provided manually reconstructed Data Flow Diagrams for 17 Java Spring microservice projects. Since the effort is manual, one of the inclusion criteria was that a project is small enough for manual analysis of source code to be feasible. Later the authors published the \texttt{Code2DFD} tool \cite{Code2DFD23}, which is used for automatic DFD generation and is the basis of this work. 

In their paper, d'Aragona et al.~\cite{amoroso2024dataset} expanded upon \cite{imranur2019curated} and performed a more thorough and systematic effort to obtain OSS Microservice systems by querying the World of Code dataset and filtering the projects by their activity, size, and amount of contributors. Authors detect the microservice systems by the presence of Dockerfiles and services defined in the \texttt{docker-compose.yml}. The authors obtained 3804 projects, which they analyzed and labeled manually to confirm that they are microservice systems. In the end, the authors provide 378 projects. 

The most recent Microservice dataset is due to Yang et al.~\cite{yang2024feature}.
The authors searched GitHub for microservice systems in Java and filtered the results to include only Java Spring projects with at least 4 services, service registration, and discovery mechanisms, and several release tags. For the provided projects, authors calculated 23 different metrics applicable to Java Spring Boot systems.

\ReviewerA{As for SMs in Software Engineering},  Gil and Lalouche \cite{gil2017correlation} discovered that qualitative metrics such as \emph{maintainability} and \emph{robustness} can be predicted from correlation with size metrics such as Lines of Code and Number of Methods.
Esposito et al. \cite{esposito_early_2023} have shown that Cognitive complexity better reflects the understandability of the code than Cyclomatic complexity. Selitto et al. \cite{9825885} have assessed the impact of refactoring on the readability of code by considering code comprehension metrics.
Notably, Fontana et al. \cite{fontana2019architectural} found that there is no correlation between \emph{architectural smells} and \emph{code smells} co-occurrence. \ReviewerA{Above-mentioned SMs have also been applied in MSA context \cite{10.1145/3143434.3143443, raj2021performance, hou2024addressing}}.

Centrality metrics have been used in Software Engineering and Architecture to assess the importance of program's classes \cite{wang2012comparative} and fault localization 
 in monolithic \cite{zhu2011software} as well as microservice \cite{mariani2018localizing} systems, although in these works the authors did not compute the centrality of microservices but of code statements and KPIs, respectively. \ReviewerC{Selimi et al. \cite{7973726} attempt to cluster nodes (servers) in a \emph{micro-cloud} and consider the centrality scores of the nodes in each cluster as a potential way to assess the quality of fit in their future work.}
 
\ReviewerA{Specifically for MSA networks, Farsi \cite{farsi:hal-03825330} has provided an overview of how centrality scores are applicable to MSA. Findik dedicated his master thesis \cite{findik_2024} to exploring how different features of a dynamically reconstructed network of an MS system, such as centrality and modularity, can be used to predict response time. Engel et al. \cite{10.1007/978-3-319-92901-9_8} have conducted interviews to determine what issues MSA developers are facing and proposed metrics based on dynamically reconstructed SDGs to address them. However, when it comes to centrality, they only consider the amount of calls, i.e. degree centrality of an MS, as do most other works \cite{al2022using, cerny2023catalog}.}

\ReviewerA{ Gaidels and Kirikova \cite{gaidels2020service} authored the contribution closest to ours. In their work, the authors calculate a diverse set of centrality scores and community structures on an SDG and argue for the adoption of such network methods to study MSA. However, the authors use their own prototype MSA as the case study and do not correlate centrality scores to other metrics associated with the MSA; they only analyze the ranking of the nodes due to centrality and the statistical outliers.
}

\ReviewerA{Specifically, the authors highlight the lack of statistical validation and the use of a single prototype system as threats to the validity and generalization of conclusions.}


\section{Empirical study design}
\label{sec:method}
In this section, we describe the study's goal, following the guidelines by Wohlin et al. \cite{DBLP:books/daglib/0029933}. We report the research questions, hypotheses, and data collection and analysis methods.

\subsection{Goal, Research Questions, and Hypotheses}
We formalized the \textbf{goal} of this study as follows:

\textit{Analyze} microservice \textbf{centrality} 
\textit{for the purpose of} evaluating its \textbf{correlation} with size, complexity, quality \textbf{metrics}
\textit{in the context of}  microservice architecture.

Based on this goal, we defined two \textbf{Research Questions} (\textbf{RQs}), which serve as the focus of our investigation. 

\begin{boxC}
\textbf{RQ$_1$} Does microservice centrality correlate with size metrics?
\end{boxC}

In the context of software development and maintenance, size metrics were classically used to estimate the effort a developer would employ to modify a specific software entity \cite{ccarka2022effort, esposito2023uncovering}. For instance, Carka et al. \cite{ccarka2022effort} used size metrics to normalize effort-aware metrics to improve their ranking capabilities in defect prediction.

Gil and Lalouche \cite{gil2017correlation} have proven empirically that more qualitative metrics such as \emph{maintainability} and \emph{robustness} can be predicted from correlation with size metrics such as Lines of Code and Number of Methods.

We aim to see whether size metrics and CMs are related. If centrality cannot be predicted from other metrics, we can argue that CMs could provide a novel perspective to the analysis of microservice architecture and its degradation. Hence, we conjecture the following null and alternative hypotheses:

\begin{itemize}
\item \emph{\textbf{H$_{01}$}: There is no correlation between microservice centrality and size metrics.}
\end{itemize}

\begin{itemize}
\item \emph{\textbf{H$_{11}$}: Microservice centrality and size metrics are correlated.}
\end{itemize}

Increasing the size of the codebase can correlate with its understandability and increase its \emph{complexity}. Thus we ask:

\begin{boxC}
\textbf{RQ$_2$} Does microservice centrality correlate with their code complexity?
\end{boxC}

Complexity metrics summarize how easy it is to understand and manage code. For example, Esposito et al. \cite{esposito_early_2023} have performed an empirical investigation to establish that Cognitive complexity better reflects the understandability of the code than Cyclomatic complexity. Selitto et al. \cite{9825885} have assessed the impact of refactoring on the readability of code by considering code comprehension metrics.

Similarly, we aim to see how the complexity and comprehension of the microservice's code are correlated by its importance in the network, i.e., the centrality score.

We conjecture the following hypotheses:

\begin{itemize}
\item \emph{\textbf{H$_{02}$}: There is no correlation between microservice centrality and its code complexity.}
\end{itemize}

\begin{itemize}
\item \emph{\textbf{H$_{12}$}: Microservice centrality and code complexity are correlated.}
\end{itemize}

The inability to understand the codebase due to its complexity can lead to bad practices, quick solutions, and, thus, technical debt. To address technical debt in microservices, we ask:

\begin{boxC}
\textbf{RQ$_3$} Does microservice centrality correlate with code quality and technical debt?
\end{boxC}

Technical debt describes the amount of work required in a project to solve existing issues, improve the quality of code, and raise it to a specific standard. Significant research effort has been dedicated to analyzing quality metrics such as technical debt \cite{robredo2024evaluating, gigante2023resolving, d2023technical, lenarduzzi2021technical, lenarduzzi2020long}.

Notably, Fontana et al. \cite{fontana2019architectural} investigated whether \emph{architectural smells} are co-occurring with \emph{code smells}, and found that there is no correlation.


We consider the same problem from a different angle - we aim to see how quality metrics are connected to the centrality of a microservice in the network, i.e., its importance in the architecture.

We conjecture the following hypotheses:

\begin{itemize}
\item \emph{\textbf{H$_{03}$}: There is no correlation between microservice centrality and quality metrics}
\end{itemize}

\begin{itemize}
\item \emph{\textbf{H$_{13}$}: Microservice centrality and quality metrics are correlated.}
\end{itemize}
    
\subsection{Study context}
The project selection process focused on GitHub repositories from various authors. 

GitHub is the most popular online repository of OSS projects\cite{esposito2023uncovering}. As explained in Section \ref{sec:related}, to our knowledge, only four datasets provide OSS Microservice (MS) projects, and they all focus on GitHub. We use the following datasets:
\begin{itemize}
    \item Imranur et al. \cite{imranur2019curated}: We consider all 20 projects.\footnote{\url{https://github.com/clowee/MicroserviceDataset}} 
    \item Schneider et al. \cite{schneider2023microsecend}: We consider all 17 projects.\footnote{\url{https://tuhh-softsec.github.io/microSecEnD/}}
    \item D'Aragona et al. \cite{amoroso2024dataset}: Since the dataset includes different kinds of projects utilizing all possible technologies, we initially filter only Java and Java Spring projects that are systems and not frameworks or libraries and do not consider the rest. We get 44 projects from this dataset.\footnote{\url{https://figshare.com/articles/dataset/Microservices_Dataset_-_Filtered_version/24722232?file=44488199}}
    \item Yang et al. \cite{yang2024feature}: We use all 55 projects from this dataset.\footnote{\url{https://github.com/yang66-hash/microservice-catalog}}
\end{itemize}

From the four considered datasets, we obtain 136 GitHub repositories, out of which 125 are unique repositories, and 11 are duplicates: no projects appear in all 4 datasets; 1 project appears in 3 datasets, missing only from \cite{amoroso2024dataset}, and 7 projects appear in 2 datasets. Of the rest,  3 are not available anymore, 3 are fork-duplicates: forks of other projects in the merged dataset, from which we consider only the most advanced in terms of git history, and 7 are not primarily written in Java (from \cite{amoroso2024dataset} only Java Spring projects were included initially).

The resulting selection contains a total of 112 OSS Java Spring projects.

\subsection{Data collection}
\label{subsec:da}
In this section, we describe the data collection procedures and results.

\begin{figure*}
    \centering
    \includegraphics[width=0.9 \linewidth]{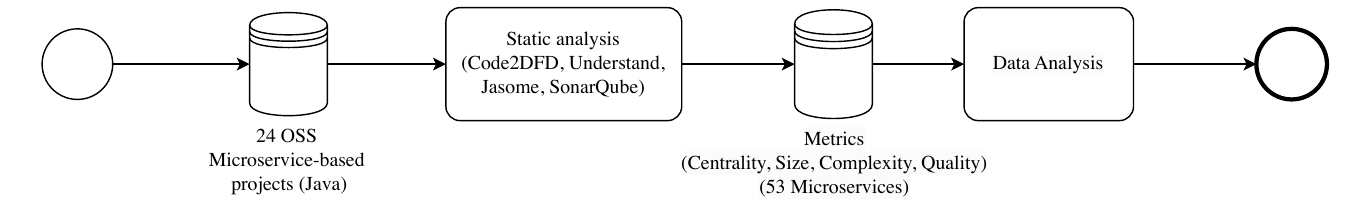}
    \caption{Data collection and analysis process.}
    \label{fig:collection}
\end{figure*}

Figure\ref{fig:collection} uses BPMN notation to present the study setup and data collection process.
Business Process Model and Notation (BPMN) 2.0 is the business processes modeling standard \cite{bpmn2_specification}. It provides a graphical notation to intuitively describe a process to technical users and business stakeholders for clear communication and collaboration. BPMN 2.0 defines various elements comprising events, tasks, gateways, and flows in precision representing complicated workflows and decision points \cite{bpmn2_specification}.

\subsubsection{Reconstructing the architecture}
We selected \texttt{Code2DFD} tool for architectural reconstruction because it proved to be the most generalizable and reliable of the tools identified in \cite{bakhtin2023tools, schneider2024comparisonrr, schneider2024comparison} during our pilots to build the dataset for this work.

We run the \texttt{Code2DFD} tool on all 112 projects on the most recent release tag, if available, otherwise on the most recent commit. In case there were more than 100 commits on \texttt{main} branch since the last release, we ran the reconstruction on the latest commit. \ReviewerB{Our online appendix and replication package report the summary information about the used projects as well the version reconstructed \cite{anonymous_2024_replication}}.

\texttt{Code2DFD} provides the reconstruction output as a \emph{directed} network of \emph{components} (split into two types - \emph{microservices} and \emph{external components}) connected via \emph{information flows} (multiple edges are not permitted, all possible properties of a flow are properties of a single edge).

We manually examine all the reconstructed networks to understand the quality and reliability of the reconstruction.
We are left with 24 projects for which the tool provided useful output, while during the analysis of 11 projects, the tool crashed; for 26 projects, only the connections of microservices to \emph{infrastructural components}, such as service discovery, API gateway, configuration server, message bus were discovered; for 3 projects, only connections of services to their respective \emph{databases} were identified; for 10 projects, only services and \emph{no connections} were identified; for 2 projects, the tool executed successfully but provided \emph{empty} output.

For the remaining 36 projects we obtained the networks in the output. However, we cannot use them further due to the following considerations:

\ReviewerB{Networks reconstructed by \texttt{Code2DFD} are \emph{simple} by design \cite{schneider2023automatic}. The network of a microservice system should be \emph{connected}. A simple and connected acyclic network is a \emph{tree} \cite{mathworld_tree}. A tree with $N$ nodes has $N-1$ edges \cite{mathworld_tree}. A network with cyclic dependencies would have even more edges.}

For the 36 projects mentioned above, the reconstruction output contains significantly fewer connections than components.
Since \texttt{Code2DFD} gets the components from Docker Compose, Maven, and Gradle definitions and then infers connections between them by analyzing the project's source code, we consider the detection of components to be more reliable than connections. So, for these 36 projects, we observed a high rate of false negatives (FNs) in connection identification.
The absence of many connections will make the centrality scores unreliable. Thus, we have to exclude these 36 projects from the analysis.

\begin{figure*}
    \centering
    \includegraphics[width=0.80\linewidth]{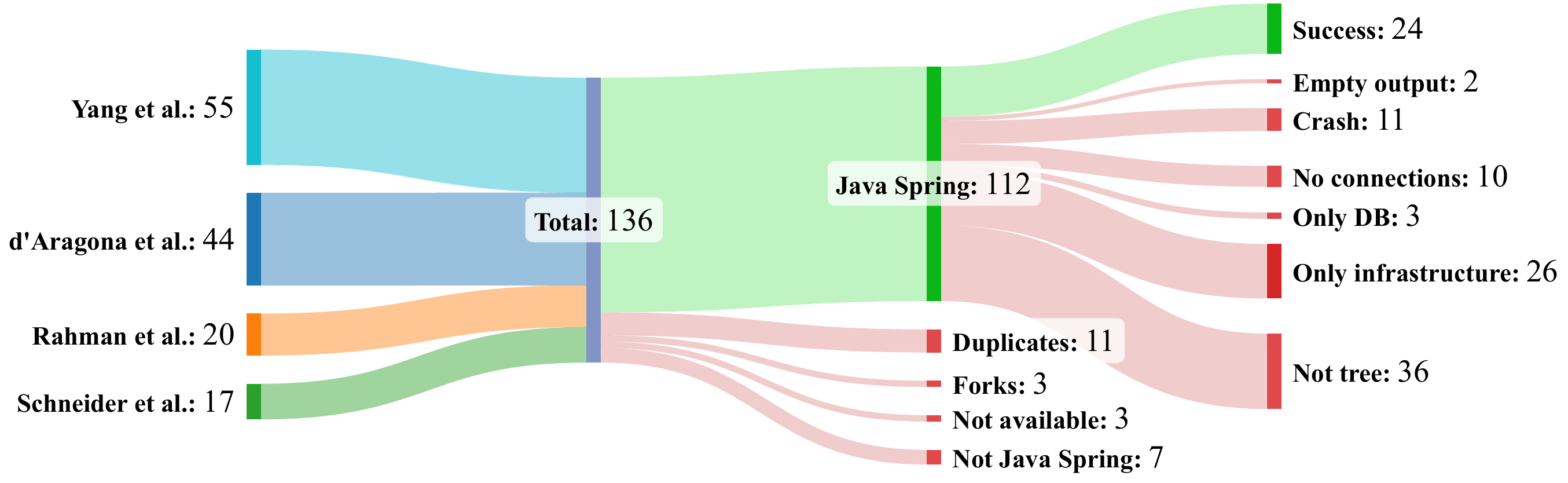}
    \caption{Different categories of collected projects.}
    \label{fig:dataset}
\end{figure*}

After the manual validation, we are left with 24 projects, which we obtained from \texttt{Code2DFD} output reliable enough to answer the RQs.
Figure \ref{fig:dataset} shows the relationships between different categories of projects. Information on all the discovered projects is available in the replication package \cite{anonymous_2024_replication}.

\subsubsection{Calculating the centrality metrics}
To properly calculate microservice centrality, we pre-process the network in the following ways:

\paragraph{Keep only the greatest weakly connected component} Since the reconstruction is imperfect and suffers from FN for the connections, the reconstructed networks contain several disconnected components (sometimes consisting of a single disconnected service). Since the centrality analysis relies on connections to determine the importance of the nodes, we should not consider these smaller components, since their scores would be inaccurate anyway. We detect the greatest weakly connected component and remove the rest of the nodes.
\paragraph{Remove databases connected to only one service} 
\ReviewerB{
\texttt{Code2DFD} tool detects database instances in addition to services. 
We remove all nodes containing \texttt{mysql}, \texttt{mongo}, or \texttt{database} in the name and connect to only one other node. In this way, we filter only the databases that fulfill the \emph{one DB per service} pattern \cite{newman_database_per_service, baeldung_microservices_db_design} and do not negatively affect the architecture. Databases that are accessed from several services in violation of best practices would be kept in the network and affect the calculation of centrality scores. Still, we would not assess the DB nodes directly since we cannot calculate SMs for them, as their code is not part of the studied projects. In our case, all databases were filtered from data, indicating adherence to the pattern.}
We calculate the centrality scores using the implementations provided by \texttt{networkx} Python library.
\texttt{networkx} is widely used for network analysis tasks; thus, providing analysis through this tool enables easier adoption by the network science community.

\subsubsection{Calculating the size metrics}

To compute the size metrics, we take the source code of all the projects and process it with the \texttt{Understand} tool. The tool provides a command line interface\footnote{\url{https://support.scitools.com/support/solutions/articles/70000582798-using-understand-from-the-command-line-with-und}}, enabling it to be effectively executed in batch on all the projects and get all the data.

Our script utilizes the CLI tool to automatically create an \texttt{Understand} project, configure the tool to query all possible metrics, perform the analysis of the created project, and save all the metrics in a CSV file. After that, we filter only the metrics given on the \texttt{Package} level.

\texttt{Code2DFD} identifies the microservices by the names of Docker containers or Maven/Gradle projects. Each studied project is a single repository where all the microservices are stored, while the \texttt{Understand} tool gives metrics for the packages declared in the Java source. For these reasons, we need to match both metrics with the microservices to compare the metrics and compute the correlation. We achieve this by manually going through all the Java packages identified by \texttt{Understand} and mapping them to component names extracted by \texttt{Code2DFD}. For example, \texttt{Basilisk} project contains the package \texttt{org.dicegroup.basilisk.benchmarkService}, while the architectural reconstruction contains \texttt{benchmark-service} microservice. Thus, this package can be mapped to this service.

After reviewing all the data, we matched 635 packages to 113 services.

Since a package can contain sub-packages, for which metrics are reported separately, we keep only metrics given by absolute values, i.e., that have \emph{Count} in their name and not \emph{Avg} or \emph{Ratio}, and sum them for all the sub-packages of the given services. From the example above, package metrics for all sub-packages of \texttt{org.dicegroup.basilisk.benchmarkService.*} are summed up and reported as the size metrics of \texttt{benchmarkService}.

\subsubsection{Calculating the complexity metrics}

We analyze all our selected projects using \texttt{Jasome} and export all possible metrics for all levels. Since the output contains package information, we can aggregate the data from packages into microservice similar to \texttt{Understand}. We map all packages to microservices using the same mapping and aggregate all metrics for a given microservice in three ways: \emph{sum}, \emph{average}, and \emph{max}, wherever applicable, i.e. if the metric for a class is a ratio of other metrics, it does not make sense to use \emph{sum} to aggregate across the packages, so we only use \emph{avg} and \emph{max}.

\subsubsection{Calculating the quality metrics}

We set up each of the analyzed repositories as a SonarQube project and ran the analysis on source code only, without compilation. We use the SonarQube API to query the quality metrics of all projects per each directory in the source code. We can map the directory path to the package since, by Java standards, the directory path should replicate the project name. For example, \texttt{org.dicegroup.basilisk. benchmarkService} package would be found in \texttt{src/main/java /org/dicegroup/basilisk/benchmarkService} directory. We mapped the paths to packages and then packages to microservices using the same package map as defined for \texttt{Understand}. Most of the metrics are aggregated for each microservice using \emph{sum}, except rating metrics, which are given on a scale of 1-5, aggregated using \emph{avg} and \emph{max}.

After combining all the data and leaving only the microservices for which all metrics are available, we have a data set of 53 services from 13 systems, for which we calculated \ReviewerA{23 Understand metrics, 53 Jasome metrics, 19 SonarQube metrics, and 11 centrality metrics}, which we used to answer our research questions.

\begin{table*}
\centering
\footnotesize
\caption{Spearman's $\rho$ interpretation.}
\label{tab:spearmaninterpretation}
\begin{tabular}{l|ccccccccccc}
\hline
               & \multicolumn{1}{c}{Perfect} & \multicolumn{3}{c}{Strong} & \multicolumn{3}{c}{Moderate} & \multicolumn{3}{c}{Weak} & \multicolumn{1}{c}{Zero} \\ \hline
\multirow{2}{*}{$\rho$ } & + 1                         & + 0.9   & +0.8    & + 0.7  & +0.6     & + 0.5   & + 0.4   & + 0.3  & + 0.2  & + 0.1  & 0                        \\
                             & - 1                         & - 0.9   & - 0.8   & - 0.7  & - 0.6    & - 0.5   & - 0.4   & - 0.3  & - 0.2  & - 0.1  & 0                        \\ \hline
\end{tabular}%
\end{table*}
\subsection{Data analysis}
This section presents how we analyzed the collected data to answer the research questions.

To answer our RQs, we have computed the centrality metrics from the architecture graph obtained by \texttt{Code2DFD}. Likewise, we gathered size metrics from \texttt{Understand} and \texttt{Jasome} (RQ$_1$), complexity metrics from \texttt{Jasome} and \texttt{SonarQube} (RQ$_2$), and quality metrics from \texttt{SonarQube} (RQ$_3$).

We must assess the metrics' distribution before choosing the proper correlation test \cite{falessi_enhancing_2023}.
To test for normality distribution of the data, we conjectured the following hypothesis:

\begin{itemize}
\item \emph{\textbf{H$_{1\mathcal{N}}$}: The gathered metrics are \emph{not} normally distributed.}
\end{itemize}

The corresponding null hypothesis is:

\begin{itemize}
\item \emph{\textbf{H$_{0\mathcal{N}}$}: The gathered metrics are normally distributed.}
\end{itemize}

We tested \emph{\textbf{H$_{0\mathcal{N}}$}} using Anderson-Darling (AD) test \cite{anderson1952asymptotic}. AD tests whether data points are sampled from a specific probability distribution, in this case, a normal distribution. AD evaluates the differences between the cumulative distribution function of the observed data and the hypothesized distribution \cite{anderson1952asymptotic}. According to Mishra et al., \cite{mishra2019descriptive}, the Shapiro-Wilk (SW) test \cite{shapiro1965analysis} would be more appropriate when using smaller datasets with less than 50 samples. However, our dataset is big enough to use AD, the more powerful statistical tool for detecting most departures from normality \cite{stephens1974edf, stephens2017tests}.
\ReviewerA{As mentioned in Section \ref{subsec:da}, we aggregated metrics across packages and classes to create a single value for a microservice. We computed sums of 85 metrics, averages of 45 metrics, and maxima of 14 metrics as aggregation methods. Also, ratios of counts of protected vs public and private vs public methods were added as size metrics. This additional computation brought the total number of unique metrics we analyzed to \textbf{155}.}

Results of the statistical test allowed us to \textbf{reject the null hypothesis} \emph{\textbf{H$_{0\mathcal{N}}$}} in 146 out of 155 cases, thus asserting that most of the gathered \textbf{metrics are not normally distributed}.
Therefore we choose to \textbf{exclude} metrics for which the p-value was high from the further analysis, namely: average values of \texttt{IOVars}, \texttt{Ma}, \texttt{Md}, \texttt{Di}, \texttt{NMA}, \texttt{NM}, \texttt{NOL}, \texttt{PMd}, \texttt{NPM} size and complexity metrics from the \texttt{Jasome} tool\footnote{\url{https://github.com/rodhilton/jasome/blob/master/README.md}}.

Since the data are not normally distributed, we choose Spearman's $\rho$ \cite{spearman_rho} to test the correlation of different software metrics with the centrality metrics, instead of \ReviewerB{Pearson's test}, which requires normally distributed data\cite{pearson1895vii}. The non-parametric test evaluates the monotonic relationship between variables, assessing whether a change in one variable leads to another, either in the same direction (positive correlation) or different (negative correlation). \ReviewerA{To interpret $\rho$ values, we adopt Dancey and Reidy interpretation \cite{dancey2007statistics} as shown on Table \ref{tab:spearmaninterpretation}}.

\ReviewerA{Finally, we adopted a stricter significance threshold of  $\alpha = 0.01$  to balance the trade-off between controlling Type I and Type II errors and maintaining statistical power while testing 155 correlations. Due to the volume of pairwise correlation tests, the Bonferroni correction would result in a stricter threshold ($\alpha^{\prime} = 0.00032258$), deemed overly conservative for an exploratory study \cite{abac.12172, esposito_correlation_2024, ccarka2022effort} leading to a substantial increase in the likelihood of Type II errors, i.e., potentially obscuring weaker but meaningful correlations. }

\section{Results}
\label{sec:result}
\begin{figure*}
    \includegraphics[width=\linewidth]{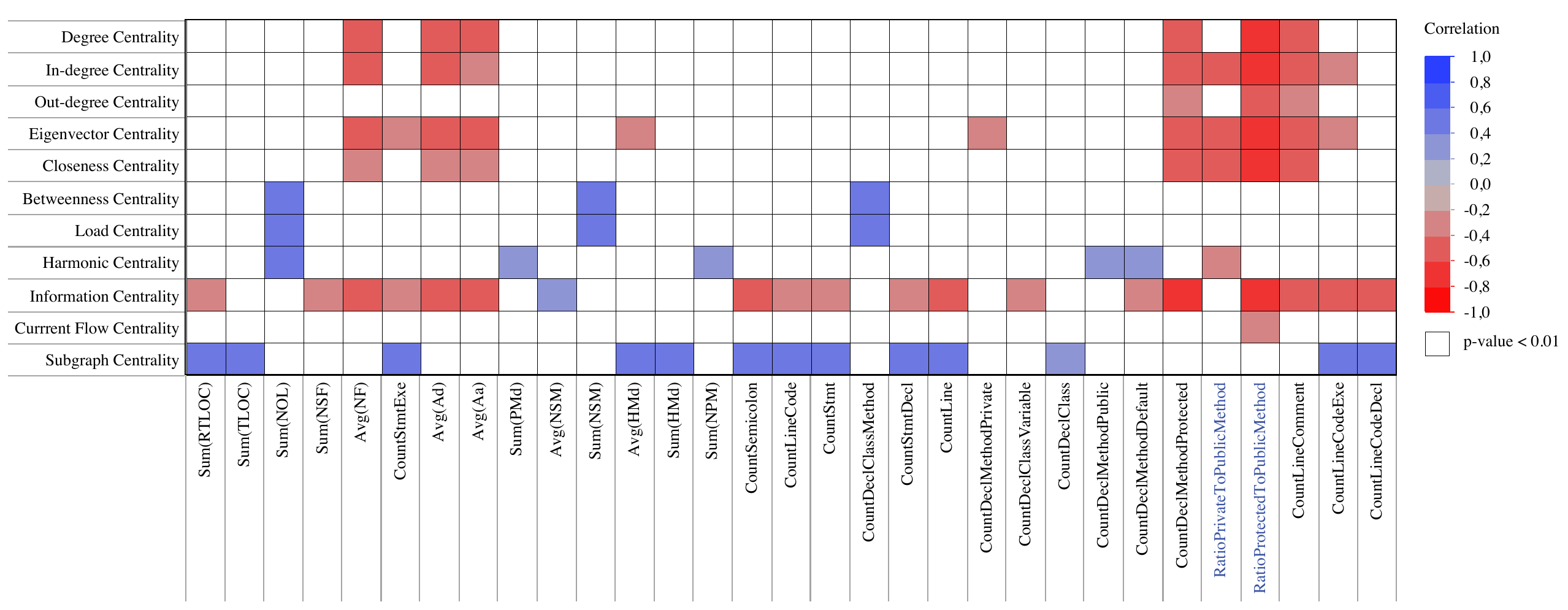}
    \caption{Heat-map of correlation of microservice centrality with size metrics.}
    \label{fig:size}
\end{figure*}

\begin{figure}
    \includegraphics[width=0.85\linewidth]{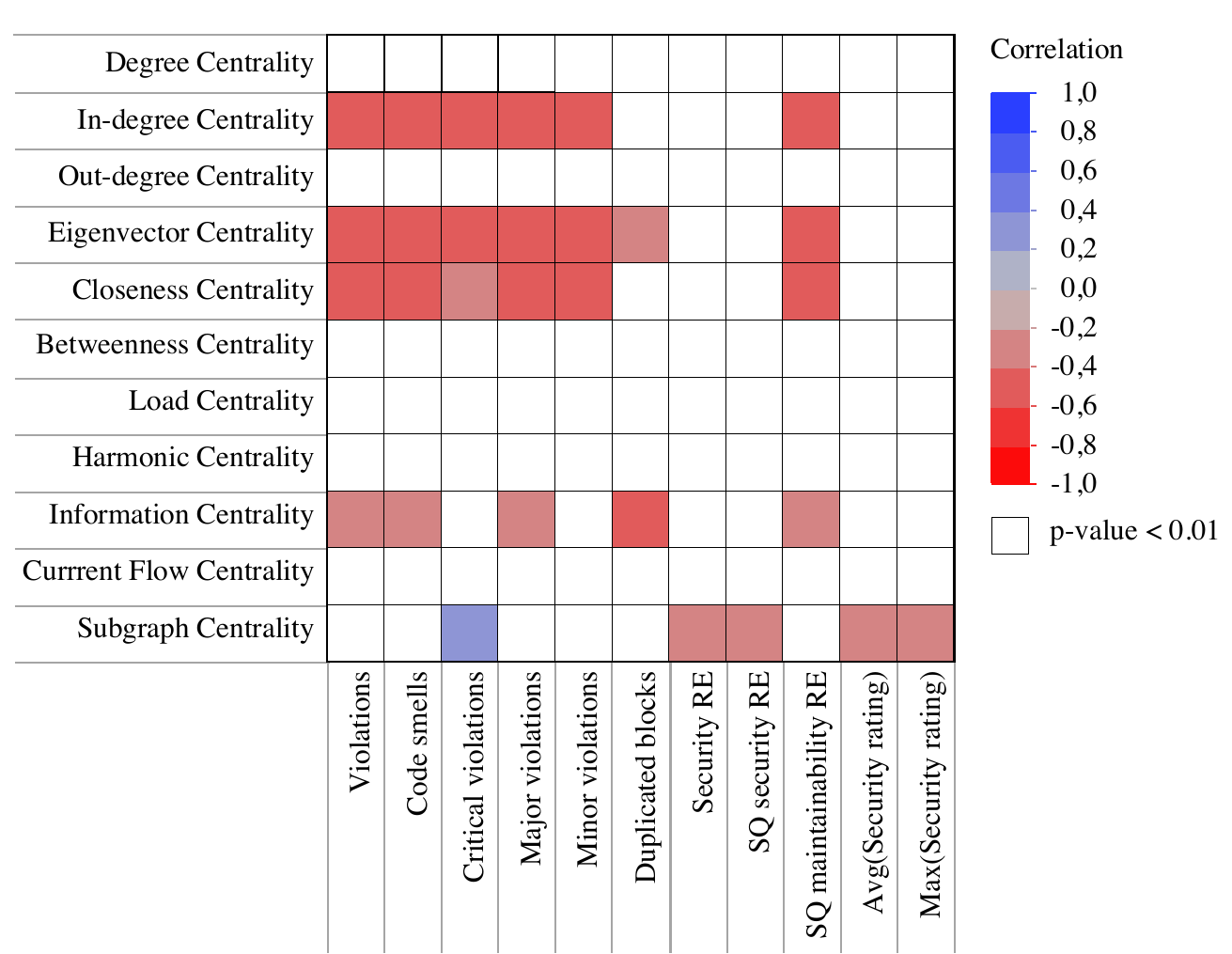}
    \caption{Heat-map of correlation of microservice centrality with quality metrics.}
    \label{fig:quality}
\end{figure}

\begin{figure*}
    \centering
    \includegraphics[width=0.95\linewidth]{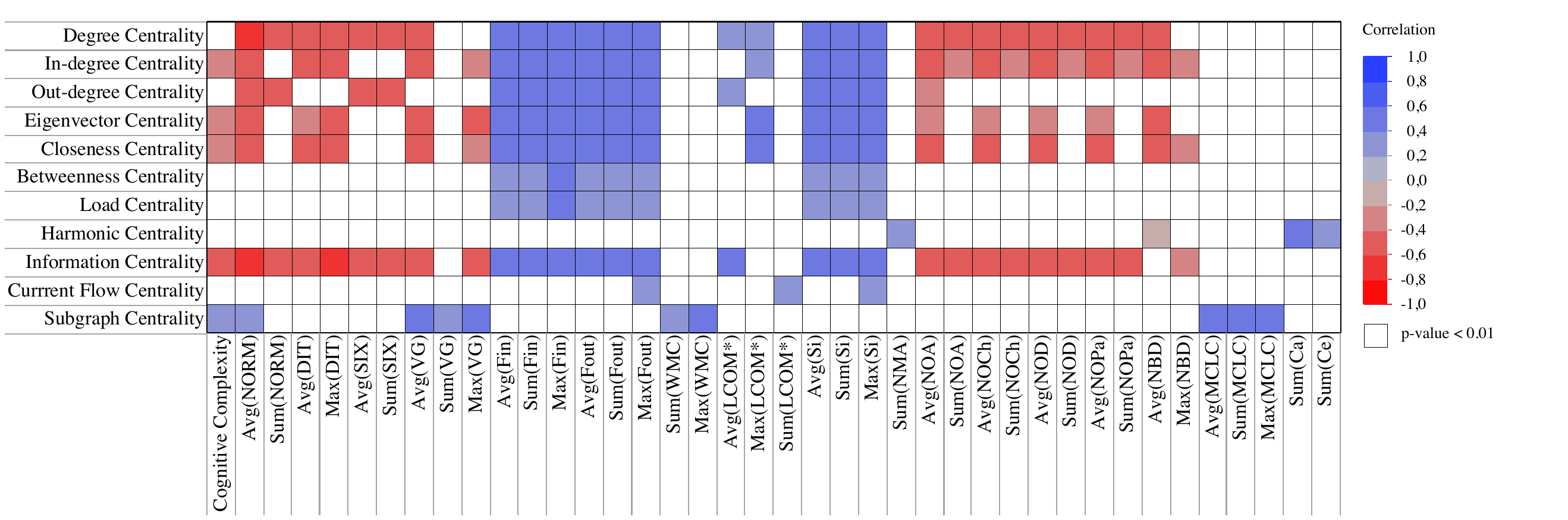}
    \caption{Heat-map of correlation of microservice centrality with complexity metrics.}
    \label{fig:complexity}
\end{figure*}

This section addresses the Research Questions by presenting the correlation analysis outcomes. The results of correlation analysis of size, complexity, and quality metrics with centrality are reported in Figures \ref{fig:size}, \ref{fig:quality},  and \ref{fig:complexity} respectively.


\subsection{Correlation with size metrics (RQ$_1$)}
We computed the Spearman's $\rho$ correlation of complexity metrics calculated by \texttt{Jasome} and \texttt{Understand} with microservice centrality.
Figure \ref{fig:size} shows the $\rho$ value between microservice centrality and complexity metrics for statistically significant correlations and indicates for which pairs the correlation is not statistically significant.

According to Figure \ref{fig:size}, \ReviewerA{11 centrality and 31 size metrics} require 341 statistical tests. We can reject the null hypothesis \emph{\textbf{H$_{01}$}} only in \ReviewerA{80 out of 341 cases}. Therefore, the following analysis focuses only on the statistically significantly correlated pairs.
Microservice centrality in the SDG, constructed based on public calls made between the services, is moderately positively correlated with the amount of Public Methods declared in its classes (NPM, PMd, CountDeclMethodPublic).
\ReviewerA{Moreover, central services tend to have low ratios of private-to-public and protected-to-public methods. Some services exclusively contain public methods, resulting in a ratio of zero.}
However, the centrality is negatively correlated with the amount of Private or Protected classes (CountDeclMethodPrivate, CountDeclMethodProtected) and the size of the classes as given by CountLineCodeExec, CountLineComment, number of attributes (NF), defined attributes (Ad), and total attributes (Aa).

We can summarize the observations as follows: \textbf{Number of exposed public methods \ReviewerA{correlates with} the centrality of a microservice in the SDG}.

\subsection{Correlation with complexity metrics (RQ$_2$)}
We computed the Spearman's $\rho$ correlation of complexity metrics calculated by \texttt{Jasome} and \texttt{SonarQube} with microservice centrality.
Figure \ref{fig:complexity} shows the $\rho$ value between microservice centrality and complexity metrics for statistically significant correlations, as well as indicates for which pairs the correlation is not statistically significant. According to Figure \ref{fig:complexity}, 11 centrality and 40 complexity metrics require 440 statistical tests. We can reject the null hypothesis \emph{\textbf{H$_{02}$}} only in 173 out of 440 cases. Therefore, the following analysis focuses only on the statistically significantly correlated pairs.

Metrics related to \emph{method} calls ($F_{in}$, $F_{out}$, $Si = F_{out}^2$) are strongly positively correlated with microservice centrality in the SDG network, which is constructed based on service calls.

Object-oriented metrics related to inheritance - number of children (descendants), number of parents (ancestors), depth of inheritance tree, overridden methods - are weak-to-moderately negatively correlated with microservice centrality.

Complexity metrics, such as McCabe's complexity, Cyclomatic complexity, and Cognitive complexity are moderately negatively correlated with centrality.
Therefore, we observe that \textbf{microservice centrality is related to the complexity and inheritance structure of its classes}.

\subsection{Correlation with quality metrics (RQ$_3$)}

We computed the Spearman's $\rho$ correlation of quality metrics from \texttt{SonarQube} with microservice centrality.
Figure \ref{fig:quality} shows the $\rho$ value between microservice centrality and quality metrics for statistically significant correlations, as well as indicates for which pairs the correlation is not statistically significant.
According to Figure \ref{fig:quality}, 11 centrality metrics and 11 quality metrics require 121 statistical tests. We can reject the null hypothesis \emph{\textbf{H$_{03}$}} only in 29 out of 121 cases. Therefore, the following analysis focuses only on the statistically significantly correlated pairs.

Quality metrics, such as the number of violations, smells, and software quality maintainability remediation efforts are strongly negatively correlated with several centrality scores. This indicates that the more important the service is in the SDG, the fewer violations of all kinds it has, according to \texttt{SonarQube}.


\ReviewerA{Similarly, security aspects, such as maximum and average security ratings and the security remediation effort, appear to be negatively correlated with subgraph centrality. Since higher security ratings should indicate better security, a high centrality score correlates to poor security. Conversely, lower security remediation effort is considered better. Thus, central microservices seem to require relatively little effort in our data to achieve better security ratings.}

Additionally, the number of duplicated blocks is weakly negatively correlated with eigenvector and information centralities.
Our findings suggest that \textbf{centrality of microservices in the network could affect its quality and security}.

Moreover, we notice that \emph{sub-graph} centrality exhibits behavior different from other studied centralities: it is positively correlated to metrics that other centralities are not related to.

\section{Discussion}
\label{sec:discussion}

This section discusses our findings. We see that more central microservices have a lot of public methods but few private or protected ones. Also, the size of the service, as measured by attributes and lines of comments and executed code, seems smaller. It is a somewhat expected result since if a microservice follows the Unix principle of \emph{doing one thing and doing it right} \cite{bstj57-6-1899}, it might have few private methods handling the data and business logic, but many public methods exposing the APIs for CRUD operations, or different types of input data for the same operations. \ReviewerA{In fact, 24 services have no private methods, and 23 lack protected methods. However, this may be due to selection bias and might not accurately reflect systems in production.}

The negative correlation of the complexity metrics with centrality, i.e., a service fulfilling one specific purpose can have few simple methods, hence low complexity and size, further corroborates our thesis. For instance, many services implementing different business logic would call a service implementing the \texttt{OAuth2.0} protocol for resource authorization. Still, the functionality of the \texttt{OAuth} service comes down to  ``simple" management of authorization keys and tokens.
\ReviewerA{Thus, a service like OAuth would be central in the SDG because it is \emph{fundamental}, not because it is complex. We could argue that centrality is related to the \emph{semantics} of the MS, not the exact code structure.}
Moreover, this also explains why central services have low inheritance complexity, i.e., developers rarely subclass classes that implement a microservice’s business logic, and, likewise,  have those classes subclass others, to avoid duplicating functionality and coupling services~\cite{Panichella2021}, which violates best practices.

Even though services do not subclass each other directly, the negative correlation of duplicated blocks with centrality can have dual explanations - either service do not have duplication because they are indeed written with best practices in mind and implement different logic while isolating implementation details, or, in violation of best practices, they depend on a set of \texttt{common} libraries or packages, thus the duplication of code is avoided, but the services are coupled due to reliance on the same set of libraries, thus changes to the libraries due to one service might break compatibility with the other, which leads to coupled updates of both services and the libraries. We can observe this in \texttt{Basilisk}, \texttt{blog-microservices}, \texttt{scaffold-cloud}, \texttt{website} systems.

Further, we recognize that if the more central services are smaller, they would naturally contain fewer violations and code smells in the smaller codebase, which leaves less room for errors.
The behavior of \emph{subgraph centrality} seems different from other considered centrality metrics since it correlates positively with many metrics for which other centralities indicate a negative correlation. One possible explanation is as follows: subgraph centrality ignores the direction of the edges and views the graph as undirected while considering all possible cycled paths in such paths. Thus, the results might be inaccurate, since paths impossible during the actual deployment of the system are considered when computing the centrality score.
On the other hand, subgraph centrality is negatively correlated with \emph{security} aspects of quality metrics while other centralities have no statistically significant correlation with them. Since the subgraph centrality considers all possible paths to compute the score, we could interpret that as considering the wider possible \emph{attack surface} of the service, since e.g. if a connection \emph{service A to service B} has been observed statically but \emph{service B to service A} has not, service B could still call service A due to an adversarial attack at runtime, for example if a modified instance of this service is injected to the production deployment.

To summarize, \ReviewerC{among 902 computed correlations of CMs and SMs, we found a weak to moderate correlation in only 282 cases.} It appears that microservice centrality largely does not correlate with existing software metrics, with the only statistically significant correlations occurring \emph{by construction} of the problem, i.e. the centrality of microservices are derived from the statically reconstructed SDG of possible calls the service can make, and the presence and amount of these calls are the result of exposed APIs which come from public methods providing access to the internal business logic the other services rely on.

\begin{figure}
    \centering
    \includegraphics[width=0.9\linewidth]{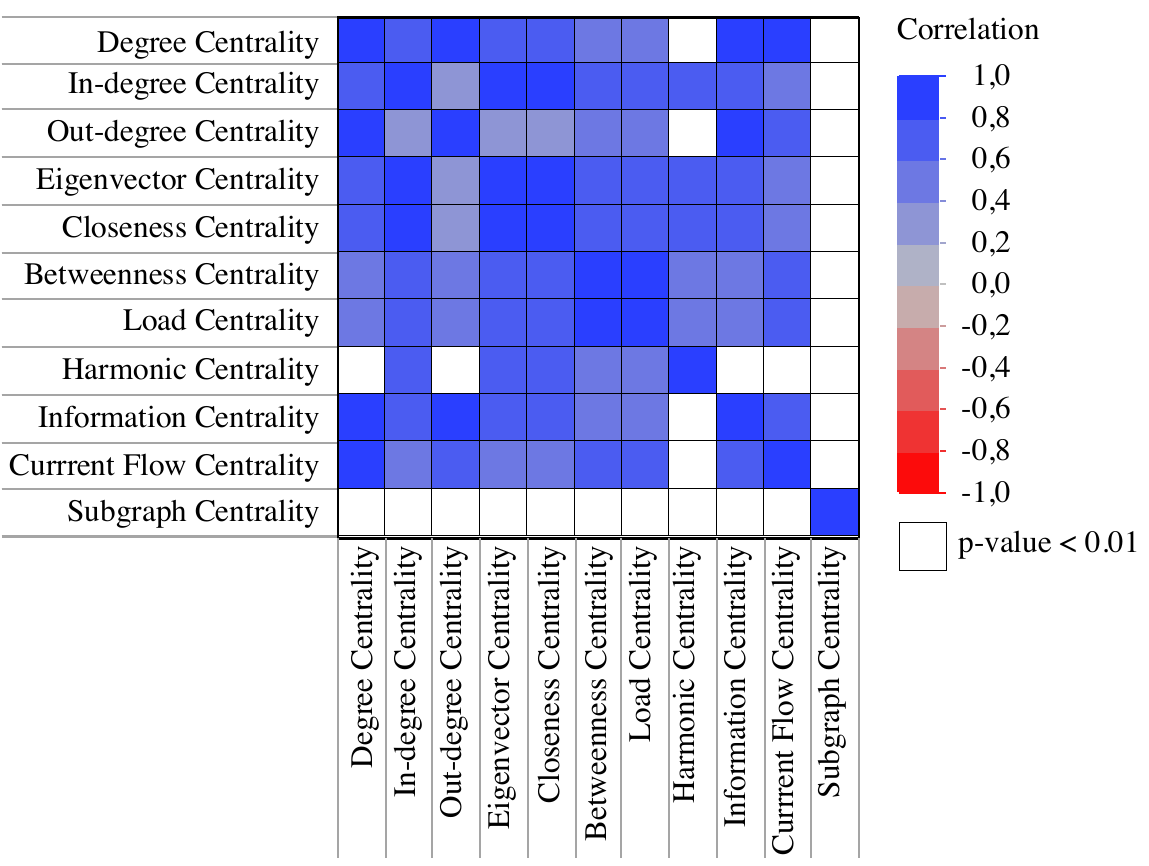}
    \caption{Heat-map of correlation between centrality metrics.}
    \label{fig:centralityheat}
\end{figure}

Thus, we propose further investigation of microservice centrality as a possible new perspective and metric for the analysis of microservices. In particular, current implementations of anti-patterns like \emph{Hub-like} as well as \emph{Nano-} and \emph{Mega-} service rely on setting the threshold on the number of calls the service has \cite{cerny2023catalog}, i.e. threshold of \emph{in-}, \emph{out-} or \emph{total} degree centrality of the service. This is an issue since the number of connections is, in principle, unbounded; for a particular system of $N$ services, it is maximally $N-1$ connections to all other services. Hence, it is hard to reason in general on how to set the thresholds for the respective anti-patterns, and current literature relies on tradition for this threshold. Adopting instead a ratio-based centrality such as betweenness centrality or normalized eigenvector centrality, which are nonetheless correlated with degree centrality
(see Figure \ref{fig:centralityheat}), would allow for a more thorough and systematic analysis of how these anti-patterns could be detected since the range of the centrality would be consistent across systems.

Finally, the unique behavior of subgraph centrality requires further investigation of its applicability towards the assessment of microservice architecture and its ability to capture other aspects of a microservice system that are not represented by metrics studied in this work, such as further focus on security aspects of the system.

\section{Threats to Validity}
\label{sec:threats}
In this section, we discuss the threats to Validity of our study, following the guidelines defined by Wohlin et al.~\cite{DBLP:books/daglib/0029933}.

\textbf{Construct Validity}. 
Our specific design choices, including our measurement process and data filtering, may impact our results. We used the centrality of microservices to measure its importance in the SDG, as is customary in network science \cite{rodrigues2019network}, and did not focus on specific interpretations of these centralities but considered many metrics simultaneously. However, different views of MSA and different assessments of microservices within it are possible. The same applies to selected size, complexity, and quality metrics - the choice is mostly motivated by the ease and freedom of use of the respective \texttt{Understand}, \texttt{Jasome} and \texttt{SonarQube} tool and the ability to easily query many metrics across many projects. \ReviewerB{Similarly, our investigation focuses on Java Spring projects due to the constraints of the \texttt{Code2DFD} tool, which allows only Java-based OSS analysis. This decision alone inherently limits the generalizability of our results to other programming languages, technologies, and frameworks. Hence, any insights gained from this study may not be generalized to microservice systems that are not based on Java or projects using other technologies. Nevertheless, Code2DFD is currently one of the most capable open-source tool for architecture reconstruction \cite{schneider2024comparison}.} \ReviewerA{Automating the reconstruction and metric computation limited the number of systems and services that were used in the final analysis due to FNs in \texttt{Code2DFD} reconstruction of connections as well as missing metrics in tools' output. However, we believe that showcasing the potential for automation and scalability in our proposed solution enhances its appeal and increases the likelihood of its adoption in the industry.}

\textbf{Internal Validity}. 
The selection of projects to be studied can bias our results since we used small OSS microservice projects, and the selection of \texttt{Code2DFD} tool restricted the choice to the Java Spring framework. To address this threat, we tried to build the most comprehensive dataset possible by merging all currently available published datasets of microservice projects \cite{imranur2019curated, schneider2023microsecend, amoroso2024dataset, yang2024feature}, and used well-established guidelines in designing our empirical study \cite{DBLP:books/daglib/0029933}. Additionally, we selected \texttt{Code2DFD} tool because it proved to be the most generalizable and reliable of the tools identified in \cite{bakhtin2023tools} during our pilots to build the dataset for this work. 

\textbf{External Validity}. 
Mining versioning systems, particularly GitHub, threaten external Validity. More specifically, since we only considered OSS Microservice projects, GitHub's user base predoGitHub comprises developers and contributors to open-source projects, potentially skewing findings towards this specific demographic. 
Moreover, the dynamic nature of GitHub, with frequent updates, forks, and merges, poses challenges in ensuring the stability and consistency of data over time. Furthermore, the accessibility of GitHub data is subject to various permissions and restrictions set by project owners, potentially hindering reproducibility and transparency in research. We addressed this issue by providing the raw data in our replication package. 
\ReviewerB{Furthermore, FNs in the reconstruction of connections with the Code2DFD tool required us to pre-process the reconstructed SDGs before computing the centralities; thus, the used networks might not fully represent the actual architecture of the system, and centrality metrics might be inaccurate. We addressed this issue by including the raw and processed SDGs in our replication package and the processing scripts, allowing the work to be replicated with other, potentially more accurately reconstructed architectures with similar or no pre-processing.}

\ReviewerB{Finally, the dataset used for our study was built based on open-source GitHub repositories. The focus on OSS constitutes a possible bias since it may not represent the average in-production microservice-based system. Larger-scale microservices or closed-source projects could exhibit behaviors and patterns not captured in our dataset; nonetheless, to our knowledge, no proper real-world, in-production MS system is currently open-sourced and available for analysis. Nevertheless, we plan in the future to perform an industry use case to test our hypothesis also in the industrial scenario.}

\textbf{Conclusion Validity}. Statistical tests threaten the conclusion's Validity and 'appropriateness of statistical tests and procedures, such as assumption violation, multiple comparisons, and Type I or Type II errors.
Spearman’s $\rho$ is a numerical measure of the strength and direction of a correlation between two variables, but due to multiple factors, this approach may hinder the conclusion’s Validity. 
In conclusion, the correlation coefficient is unreliable with a small sample size and cannot accurately represent the relationship between variables. Spearman’s correlation assumes a monotonic relationship between variables, which may lead to inaccuracies if the assumption is violated. To address this, we ensured a representative sample and verified the data distribution. Although alternative non-parametric correlation measures could be considered, we found Spearman’s $\rho$ more suitable than Kendall’s $\tau$, as it handles tied observations effectively, while Kendall’s $\tau$ relies on concordant and discordant pairs.

\section{Conclusion}
\label{sec:conclusion}
\ReviewerC{In conclusion, we analyzed 53 services and computed 155 CM and SM metrics. Among 902 computed metric correlations, we found weak to moderate correlations in 282 cases.} Our findings suggest that microservice Centrality Metrics (CM)s are weak to moderately correlated, when statistically significant, with traditional Software Metrics (SM), except when the correlations arise from the SDG's structural properties. The central services follow best practices with smaller sizes, fewer private methods, and lower complexity, implying simplicity and API-driven interaction. However, the negative correlation of duplicated blocks with centrality can indicate an opposite dynamic, i.e., dependence upon shared libraries, hence coupling of services and giving rise to maintenance problems, as we observed in several projects.
Subgraph centrality differs in behavior from the other centrality metrics assessed. It shows, on the one hand, positive correlations with metrics for which the other CMs showed a negative correlation and, on the other hand, a weak to moderate correlation with security aspects, hence possibly making it useful for the identification of architectural risks such as larger attack surfaces, nonetheless, there are possible limitations in capturing realistic system behavior by relying on undirected paths.

Our findings highlight a chance for centrality metrics to provide another perspective for microservice architecture analysis, especially in identifying the anti-patterns of Hub-like, Nano, and Mega-services. Centrality measures based on ratios, such as normalized eigenvector or betweenness centrality, can allow more systematic detection in the future by uniform thresholds over systems. 
The position of subgraph centrality in security and architectural quality deserves further studies to understand its full applicability in microservice systems.

Future works include the analysis of centrality over time, including the application of temporal network analysis techniques and the assessment of the usefulness of the CMs by practitioners.

\section*{Data Availability Statement}
The scripts, the resulting networks, and metrics are available in our Online Appendix and Replication Package\cite{anonymous_2024_replication}.

 \section*{Acknowledgment}
This work has been partially funded by projects Business Finland 6GSoft,  and Research Council of Finland MuFAno (grants n. 349487 and 349488).

\bibliographystyle{IEEEtran}
\bibliography{main}
\end{document}